\documentclass[conference,compsoc]{IEEEtran}
\ifCLASSOPTIONcompsoc
  \usepackage[nocompress]{cite}
\else
  \usepackage{cite}
\fi
\ifCLASSINFOpdf
\else
\fi
\usepackage{graphicx}
\usepackage{color}
\usepackage{hyperref}

\IEEEoverridecommandlockouts

\usepackage{tikz}
\newcommand\copyrighttext{%
  \footnotesize Copyright \textcopyright 2017 IEEE. Personal use of this material is permitted.
  Permission from IEEE must be obtained for all other uses, in any current or future 
  media, including reprinting/republishing this material for advertising or promotional 
  purposes, creating new collective works, for resale or redistribution to servers or 
  lists, or reuse of any copyrighted component of this work in other works. 
   DOI: \href{https://doi.org/10.1109/FIE.2017.8190587}{10.1109/FIE.2017.8190587}
  }
\newcommand\copyrightnotice{%
\begin{tikzpicture}[remember picture,overlay]
\node[anchor=south,yshift=10pt] at (current page.south) {\fbox{\parbox{\dimexpr\textwidth-\fboxsep-\fboxrule\relax}{\copyrighttext}}};
\end{tikzpicture}%
}

\begin{document}
%
\title{Aspects on Finding the Optimal Practical Programming Exercise for MOOCs}


\author{\IEEEauthorblockN{Ralf Teusner}
\IEEEauthorblockA{Hasso Plattner Institute\\
University of Potsdam\\
Potsdam, Brandenburg, Germany\\
Email: ralf.teusner@hpi.de}
\and
\IEEEauthorblockN{Thomas Hille}
\IEEEauthorblockA{Hasso Plattner Institute\\
University of Potsdam\\
Potsdam, Brandenburg, Germany\\
Email: thomas.hille@student.hpi.de}
\and
\IEEEauthorblockN{Christiane Hagedorn}
\IEEEauthorblockA{Hasso Plattner Institute\\
University of Potsdam\\
Potsdam, Brandenburg, Germany\\
Email: christiane.hagedorn@hpi.de}
}


%


\maketitle

\begin{abstract}
Massive Open Online Courses (MOOCs) focus on manifold subjects, ranging from social sciences over languages to technical skills, and use different means to train the respective skills.
MOOCs that are teaching programming skills aim to incorporate practical exercises into the course corpus to give students the hands-on experience necessary for understanding and mastering programming.
These exercises, apart from technical challenges, come with a series of questions to be addressed, for example: which fraction of the participants' time should they take (compared to video lectures and other course activities), which difficulty should be aimed for, how much guidance should be offered and how much repetition should be incorporated? 
The perceived difficulty of a task depends on previous knowledge, supplied hints, the required time for solving and the number of failed attempts the participant made. Furthermore, the detail and accuracy of the problem description, the restrictiveness of the applied test cases and the preparation provided specifically for a given exercise also influence the perceived difficulty of a task.
In this paper, we explore the data of three programming courses to find criteria for optimal practical programming exercises.
Based on over 3 million executions and scoring runs of participants' task submissions, we aim to deduct exercise difficulty, student patterns in approaching the tasks and potential flaws in task descriptions and preparatory videos. 
We compare our findings to in class trainings and traditional, mostly video and quiz based MOOCs.
Finally, we propose approaches and methods to improve programming courses for participants as well as instructors. 
\end{abstract}

\copyrightnotice

\IEEEpeerreviewmaketitle


\section{Introduction}
Massive Open Online Courses are, as massive and open implies, intended for broad audiences.
An optimal practical programming exercise should appeal to all participants, challenge them but also be solvable with reasonable effort in a predefined timespan.
The individual challenge thus relies on prior knowledge, and the term massive in MOOCs naturally implies that the participants taking a course bring in a wide range of prior knowledge in many areas, be it directly connected, adjacent or unrelated with the topics covered in the course.
While this spectrum of knowledge is a profitable foundation for discussions in the forum, there are also occasions where this range is hindering learning outcomes.
In wide audience settings, like forum discussions, a considerably large group of participants is reading questions written by a much smaller share of participants expressing those questions.
This automatically circumvents harmful knowledge gaps, as those participants that fit for the actual information needs will interact, while all others act as silent bystanders and eventually also learn by passively reading.
Given more narrow settings, like discussions within small groups or peer-assessments, the knowledge gap between the participants will not regulate itself.
Such a skill gap can either be conducive, as an experienced user explains concepts  that have not been understood beforehand and is lead to new thoughts by questions that have not come to mind before, or be cumbersome, as the experienced user might be bored by elementary questions.
Also, two participants being on elementary level, are most likely not best suited to find the solution to their problem without further external help.
Knowledge differences can thus either be helpful or hindering, depending on the actual setting. 
In order to steer potential outcomes and gain benefits of the knowledge differences, the prior knowledge of the participants has to be assessed.

The assessment in this case is neither really formative, meaning that it is not intended to build feedback upon or serves as a starting point for an intervention, nor is it summative, as it is not used for grading.
In order to coin our assessment, we would call it informative assessment, as it is primarily used to improve potential actions in the future. This is close to a formative assessment, however it lacks the necessity of an intervention afterwards.

For the following approaches and descriptions, we want to state that whenever we speak of an absolute skill expressed in numbers, we are aware that this numerical value can not reflect the true knowledge, experience and mastery of a topic. 
It is not intended to rank participants in kind of a high score and to display these individual values.
On the opposite, the skill levels will be used internally to optimize the learning outcomes of all participants.
The determination of suitable steps to improve course interactions itself is an open question and will be further discussed in Section \ref{ch:future-work}.
 
Assessing knowledge and skills is difficult in general. 
Companies spend huge amounts of money on elaborated approaches like headhunters and assessment centers to find right candidates for job offerings as wrong decisions come at an even higher price. 
Commercial providers like AMCAT\footnote{https://www.myamcat.com/} or others build their whole business model around the assessment of skills in various areas.
In contrast, the assessment within MOOCs does not have to offer such fine granularity and does not contain high financial risks if it is inaccurate, which makes the problem easier.
In our context, we also focus on a rather technical area, which tends to have a better graspability and expressiveness in numbers than for example communication skills.
However, the given MOOC setting also adds other difficulties.
Participants cannot be bothered with long quizzes or too excessive or too delicate questions.
As they take part in courses mostly based on intrinsic motivation and without direct career goals, posing too cumbersome hurdles will only result in participants skipping the questions or in worst case quitting the whole course.

After gaining some insights on participants' prior knowledge, the actual programming exercises come into focus.
Programming exercises that were not chosen well for the individual participant have several downsides that can result in a variety of negative effects.
Exercises being too easy will not challenge participants enough. 
While easy success might increase motivation over the first few exercises, it will increase the risk of frustration when facing exercises with a higher difficulty, as one got used to passing without effort.
Especially participants having a higher prior knowledge than that being aimed for in the target audience will be bored by exercises being too easy.
While this might seem bearable from the view of getting a heterogenous group of participants to optimize the course videos and additional course material for, in the notion of classic distance learning, this is a huge loss for MOOCs.
Having advanced practitioners or even experts of a programming language within the field of participants potentially yields tremendous benefits.
In past courses we conducted, some motivated experts helped out on various occasions, ranging from pinpointing ambiguities in videos, wording and slides over answering upcoming questions to supplying suited links or even creating additional material.
Last but not least, advanced and expert users answered forum posts in quality, length and speed simply impossible to the teaching team, as it was bound on other tasks such as technical support and additional content creation. 

The remainder of this paper is structured as follows:
Section \ref{ch:concept} shows on which data to distinguish suited from unsuited exercises. Section \ref{ch:analysis} presents the measurements we extracted from three programming courses and already draws some results. Section \ref{ch:related-work} shares related work and relates it to our approach. In the last sections, we conclude our findings, and give an outlook on our upcoming plans.


\section{Concept}\label{ch:concept}

In general, exercises being considered suitable for the advancement of a participant should either teach a new concept or deepen the understanding of a previously covered one.
As already shortly motivated, the suitability of an exercise therefore depends on  the specific (sub-)topics dealt with and on the (perceived) difficulty, composed of: 
\begin{itemize}
	\item the difficulty of the actual steps to solve the exercise,
	\item the prior knowledge of the participant,
	\item the expressiveness of the exercise description,
	\item the offered templates and hints, and
	\item additional help. 
\end{itemize}

 The perceived difficulty is most tightly correlated with the users knowledge. 
 In order to propose suitable exercises, it is therefore of vital interest to assess prior knowledge of participants.
 This paper first presents three different approaches to assess the prior knowledge, with a focus on knowledge in the field of programming and testing.

On the most abstract level, our concept builds on three pillars to approximate the actual skill:

\begin{enumerate}
  \item Ask the participants directly how they would classify their prior knowledge.
  \item Ask (multiple choice) questions of differing difficulty to determine their knowledge on an abstract level.
  \item Incorporate metrics of the ongoing course. As we focus on programming exercises, we can analyze events specific to programming, such as unit tests solved or the number of errors produced.
\end{enumerate}

Of course, these approaches should lead to similar results or even support each other.
However, there are some exceptions: participants most likely won't voice (1) that the content was too difficult - probably they will just leave the course. The resulting dropouts or stopouts \footnote{temporary dropouts with participants coming back after some time} will be reflected in the metrics (3).

The first option, simply asking, is the most trivial and might seem superficial at first.
However, pedagogical and psychological research has shown, that it is reliable at least in in-class settings~\cite{ross_reliability_2006,boud_enhancing_2013}.

The accuracy of the second option highly depends on the questions asked.
The result should allow to distinguish between candidates that have no knowledge and basic knowledge as well as advanced or even expert knowledge.
Finding a minimal set of such questions is non-trivial itself, as course instructors need to guess the actual difficulty of their survey questions with regards to the expected audience. 
Albeit testing these questions within a dry-run with a group of colleagues, friends and students in different stages, anticipating the distribution of participants enrolling into a course is hard.

The third option comes with increased effort of data acquisition and interpretation.
Furthermore, the metrics need to be based on a sufficient number of exercises solved before being treated as reliable, as outliers are especially probable during first tries and would strongly distort conclusions due to the sparse data foundation.
This postpones the availability of such evaluations on metric data to later course stages.

With regards to the reliability of the approaches, we assume that approaches (1) and (2) will be rather similar and have to be treated with caution. Not because they do not work in general, but because the self-assessment might be skewed much stronger than in a normal class setting due to a lack of fellows to compare with, and because the questions might have been too easy or too hard for our audience.
As approach (3) reflects part of the actual progress, it should be regarded as most reliable and in doubt be trusted in favor of the other approaches.
Potential metrics to incorporate for the domain of programming exercises are:

\begin{itemize}
\item the actual working times in practical programming assignments (until high score was reached), 
\item reached scores,
\item the number of runs per exercise,
\item the number of  errors per exercise,
\item typing speeds, and
\item copy \& paste events.
\end{itemize}

In order to judge the suitability of exercises already conducted in retrospective, we consider the following data to be of most interest:
First, the unique accesses of students to an exercise will reflect whether a particular exercise being much too difficult caused participants to leave the course.
Second, the reached scores will show whether an exercise was too hard to be completed in general.
Third, the required timespan to solve the exercise will give further detail on the difficulty of an exercise even if the majority of participants successfully completed the exercise.




\section{Analysis}\label{ch:analysis}

We analyzed the data of three past programming courses: an introductory course to Java in 2015 (java15), an introductory course to Python also from 2015 (python15) and an advanced course on Java JUnit testing in 2016 (junit16).

\begin{figure}
\begin{center}
    \includegraphics[width=1.0\columnwidth]{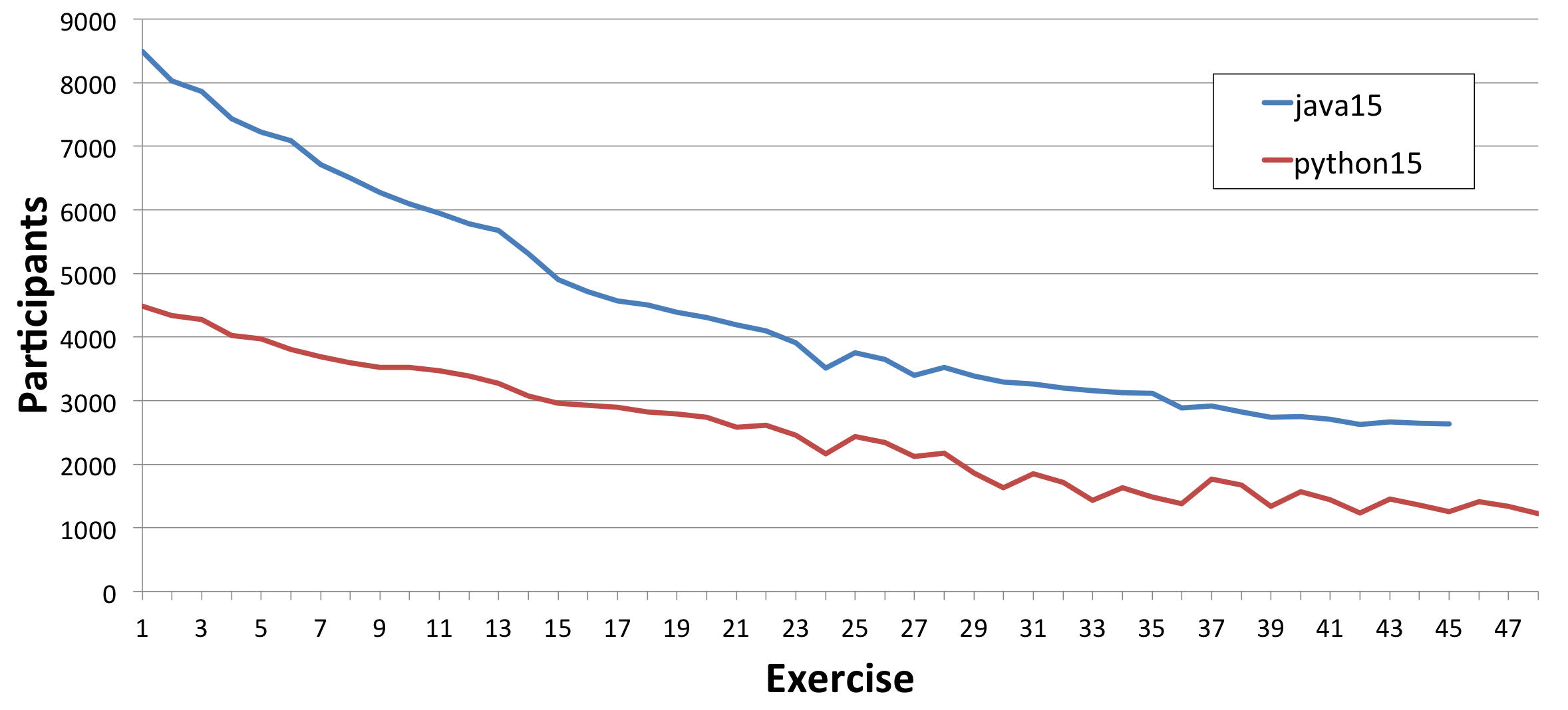}
    \caption{Number of participants per programming exercise}
    \label{pic:participants-per-task}
\end{center}
\end{figure}

\begin{figure}
\begin{center}
    \includegraphics[width=1.0\columnwidth]{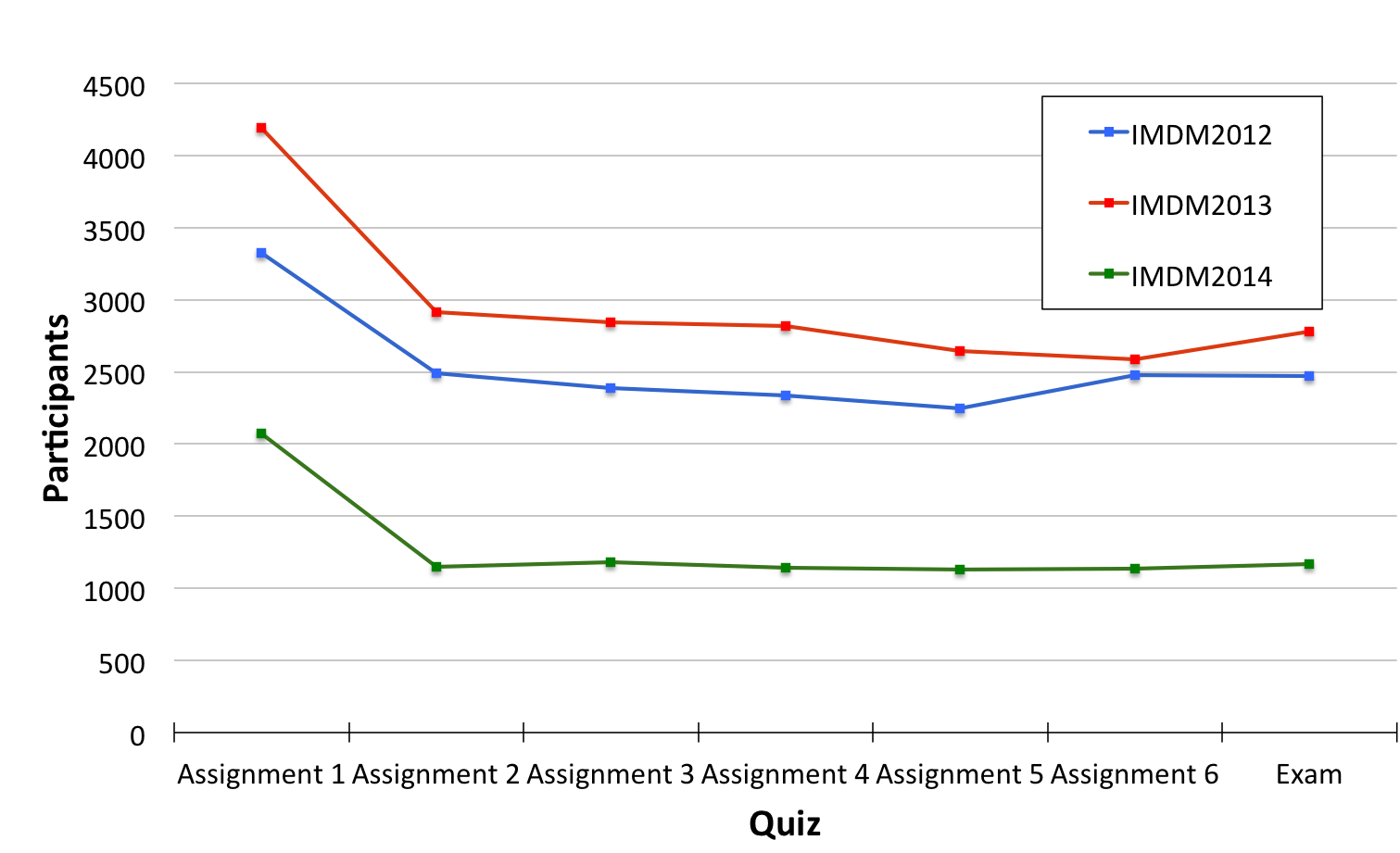}
    \caption{Number of participants per quiz in IMDM MOOCs}
    \label{pic:imdb-participants-per-quiz}
\end{center}
\end{figure}

As stated in Section \ref{ch:concept}, we start by checking the accesses on the exercises.
Since the advanced Java course only offered three exercises, we omitted it in this consideration.
From Figure \ref{pic:participants-per-task} can be seen, that participants steadily dropped out during the whole course runtimes, with a slight saturation nearing course ends.
This behaviour is different from the one we see in MOOCs that are mainly quiz based, in which we encountered a dropout of about one third of the users after the first week and having a steady user base afterwards (see Figure \ref{pic:imdb-participants-per-quiz}).

In both programming courses, the total dropout rate is about 70\% of the initial participants having started at least the first exercise.
In the Python course, a notable number of participants seemed to have noticed the underlying methodology of increasing difficulty per set of exercises for each topic and began skipping the hardest ones after the middle of the course (observable via the zigzag-pattern on the graph).

\begin{figure}
\begin{center}
    \includegraphics[width=1.0\columnwidth]{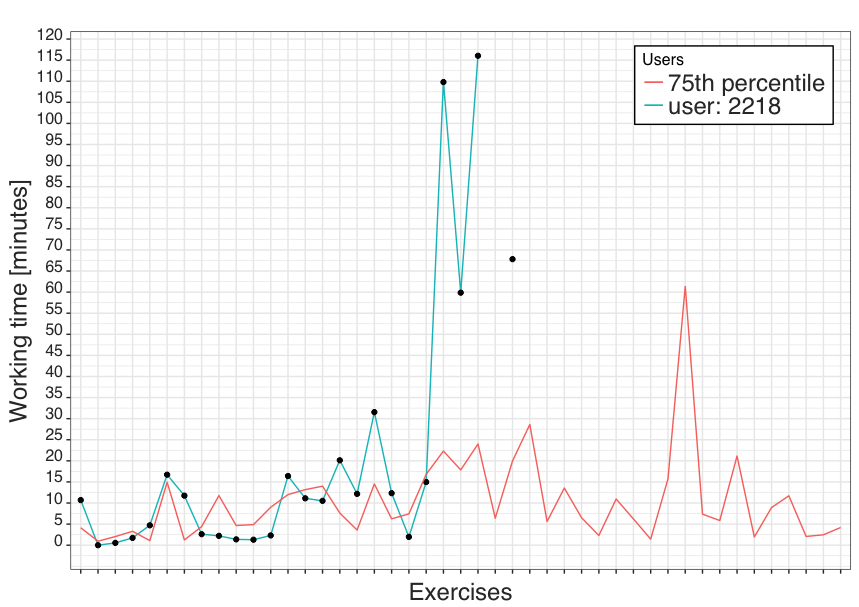}
    \caption{Working times of 75th percentile of students. Times above red line are considered to be slow.}
    \label{pic:working_time_75_percentile}
\end{center}
\end{figure}

Given these numbers, we could not deduce further insights towards the whole audience.
We therefore evaluated additional metrics that yield also helpful information for individual users, like achieved scores, number of copy and paste events and required times.
Of particular interest is the time students need to solve the assignments, which we call working time.
Most participants showed enough diligence to solve mostly all exercises completely, resulting in full scores.
Therefore the required time substitutes the score as a metric for skill to some extent.
Additionally, the required time is the base to normalize other metrics upon, such as typed characters, occurred errors or number of runs.
For all three courses, the number of runs mirrored the required time, leading us to disregard the number of runs for further analysis.
For errors, the correlation to the working time  also seems to hold, however we eventually still want to analyze the errors further, as the specific types of errors and the detection of bursts of errors is likely to yield additional insights about the kind of problems participants are struggling with. 
The analyzed working time includes the time from opening an exercise until the student reached the maximum possible score or his best score.
Due to the nature of MOOCs, where students solve their exercises at home, or wherever and whenever they want, we only consider consecutive working times with less than 10 minutes break as working time.
As we have the most data for the java15 course, the analysis of working time effects will be based on this MOOC. 
Figure \ref{pic:working_time_75_percentile} shows the working times throughout the course of a particular student (blue line) compared to the 75th percentile of all users (red line).
The exercises on the x-axis are ordered according to their occurrence in the course.
The 75th percentile on the working time (which by nature also resembles the average working time over all students with some delta) shows that most exercises were solved in 2 to 25 minutes by the majority of participants.
The only exception, a spike on the red line on the right side, happened on an exercise on polymorphism (exercise 35).
The much longer working time can be explained by two reasons in this case: the concept is probably the hardest one discussed in the course, while the length and detail of the accompanying video did not account for that accordingly, by treating this more advanced topic like any other, easier, topic.
When checking the dropout numbers after this exercise, we did not detect a noticeable increase after this exercise, which is also backed by the absence of a noticeable drop in Figure~\ref{pic:participants-per-task}.
The absence might possibly be explained by the fact that the exercise was late in the course, therefore only facing participants that already invested much into the course and therefore endured the struggles.  
Our analysis of the working times of several students showed that students who stopped out\footnote{The term stopped out in MOOCs refers to participants that show an extended absence and, therefore, potentially dropped out. As it is uncertain whether they will come back to finish the course while the course is still running, every participant that (temporarily) quits the course is considered a stopout. In case they do not return, they are additionally considered a dropout after course runtime.} often worked longer on their assignments compared to their peers.
This indicates that stopped out students often encountered problems solving the assignments resulting in longer working times.
This led to frustration and ultimately to quitting the course.
Our hypothesis is thus that students who are often slower than their peers are more prone to stop out, indicating an overextension.
Before the student under analysis  stopped out, he spent significantly more time to solve the exercises compared to his peers.
After noticing several cases like this, we put focus on the last three assignments of students before they stopped out.
For our analysis we classify a working time as slow if the student worked longer than 75\% of the other students in the assignment.

\begin{figure}
\begin{center}
    \includegraphics[width=1.0\columnwidth]{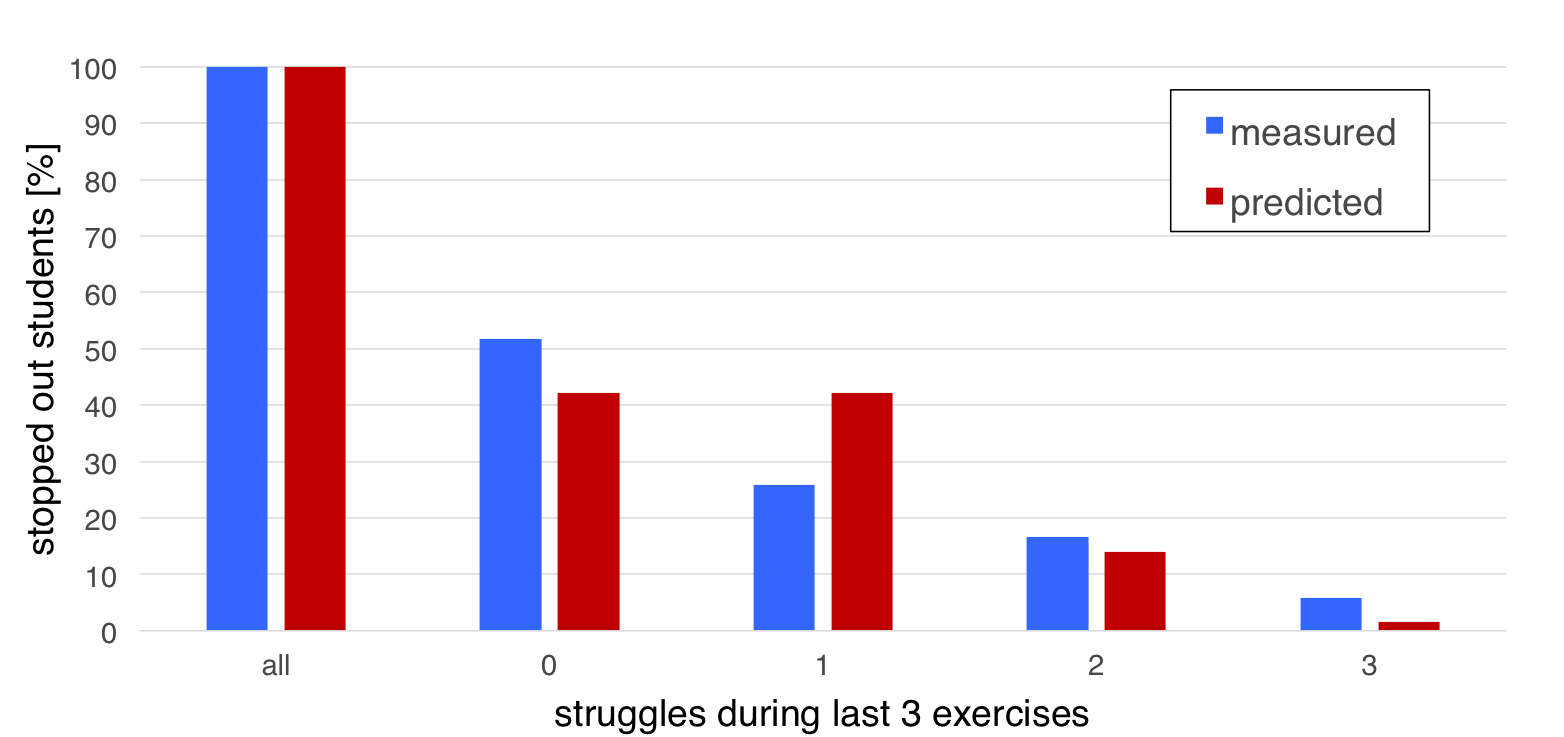}
    \caption{Percentage of stopped out students that performed slower than 75\% of average in the last 3 assignments before they stopped out}
    \label{pic:stopouts}
\end{center}
\end{figure}

In Figure \ref{pic:stopouts} we counted the stopped out students in the java15 course and calculated in how many of the last three assignments they performed slower than 75\% of all students.
The chart groups all stopped out students by the count they were slower than the top 75\% working times of all students within the last three assignments they worked on before they stopped out.
For reference, we printed the expected amount of stop outs caused by struggles.
The expected amount is built on the likelihood of a participant being in that group (for the case 0, this means he was never in the share of slowest participants, resulting in a percentage of $ 0.75 \cdot 0.75 \cdot 0.75 \approx 0.42$).  
 The uniform distribution assumes that the working times for the assignments are independent from each other. 
 From the actual measurements, we see that students stopped out without being stuck a little more often than expected (0 assignments in which they were slower).
These cases are probably caused by students having no time anymore or lost interest for some other reason.
However, it can also be seen that stopped out students who needed longer for at least one assignment tend to need more time for other assignments, too, and are therefore more likely to appear in the slower groups than to be expected by the uniform distributions.
Since they stopped out afterwards, we can assume their long working times resulted from problems with the assignments, rather than working longer for learning more or being distracted. 
As a reaction to these findings, we want to help students who are struggling before they decide to quit the course.

In the advanced Java course (junit16), we asked the participants to self-assess their skill level at the beginning of the course and also asked them several multiple choice questions with regard to programming concepts as described in Section \ref{ch:concept}.
The overall distribution of the students having completed the survey ($N=1280$) was as follows: no prior knowledge(0): 2\%, basic knowledge(1): 29\%, good knowledge(2): 42\%, very good knowledge(3): 22\% and excellent knowledge(4): 4\%.
We first checked the correlations between the scores of the multiple choice questions with the self-stated skill levels. The average score of the levels were (from skill level 0 to 4): 2.63, 3.13, 3.54, 3.7 and 3.7 out of a maximum of 4.
We see a constant rise of scores with rising skill levels.
In order to justify this impression, we applied several statistical tests.
As a correlation test, we picked Kendall's tau over Spearman's rho or the Pearson correlation, as it is less prone to outliers and we are mostly interested in the ranks. 
Kendall's tau-correlation between the skill level and the score is between 0.21 and 0.35, meaning only low to moderate correlation.

When omitting all participants that did not achieve any points and thus did not take part in the course afterwards, the correlation even shrinks to 0.14 at worst.
 Perhaps more of psychological interest is the fact, that although we asked participants to only submit their first solution, many participants re-submitted the quiz with increased scores, albeit there are no assignment points granted for this quiz.
When comparing just the first answers of every user, the correlation shows a value of 0.35, when considering only the last (mostly higher) scores, the correlation dropped to 0.21.
In the following, we only include students who achieved at least one graded point and are thus considered to haven taken part in the course, as our collected data is richer for them.
As the correlation only accounts for monotonic relations, we further considered applying Student's t-test for the group of lower skilled participants (0, 1) against the higher skilled ones (3, 4). 
However, as the variances differ too much as shown by an F-test, we decided to go for Jonckheere's trend test~\cite{jonckheere_1954}.
The trend test rejected the 0-hypothesis with the alternative hypothesis that the values are increasing (with $p = 0.0001$), stating that there is a statistically significant trend from rising skill to rising scores.

\begin{figure}
\begin{center}
    \includegraphics[width=1.0\columnwidth]{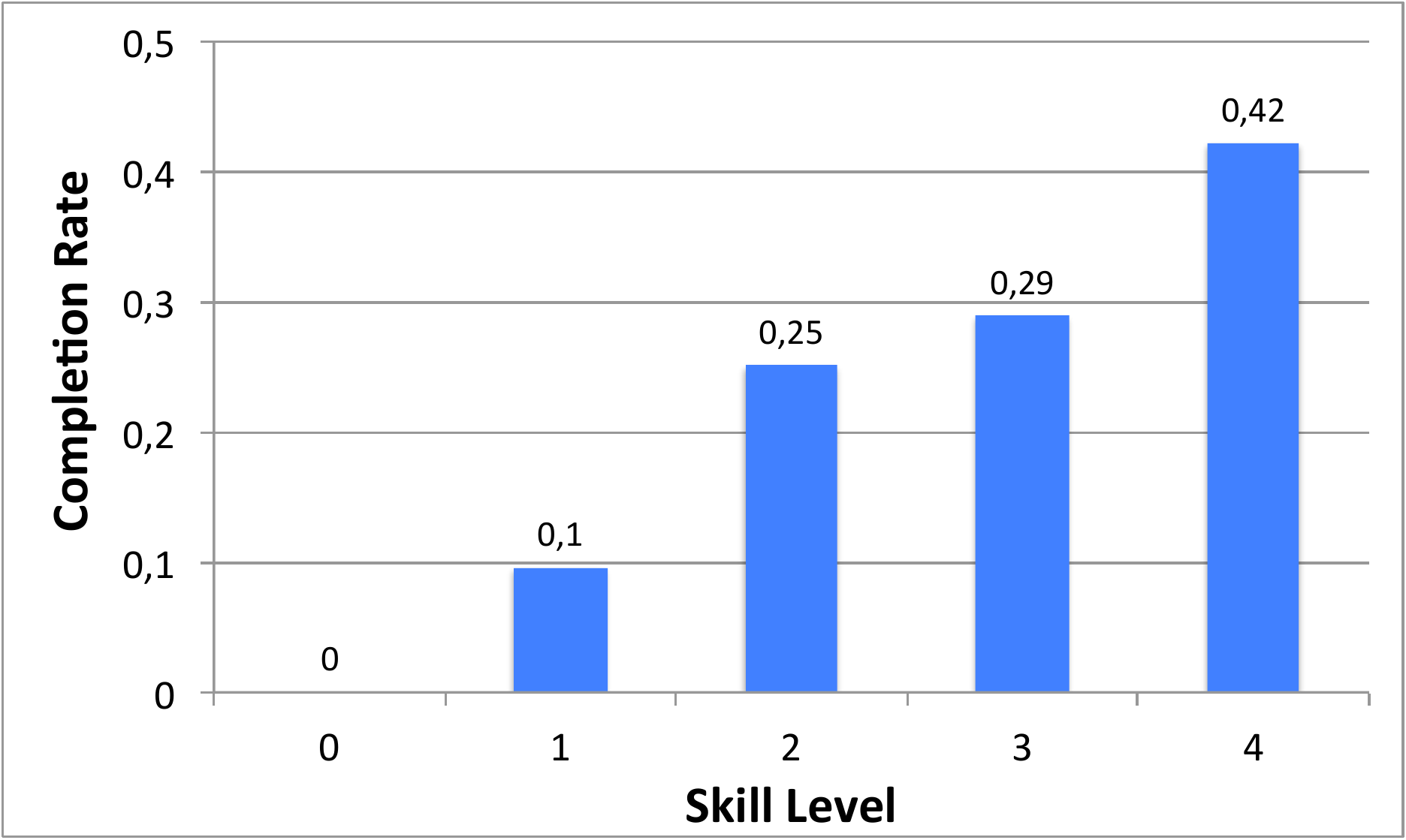}
    \caption{Likeliness of reaching the certificate based on the self-assessed prior knowledge}
    \label{pic:completePerSkillLevel}
\end{center}
\end{figure}

While we thus already gathered some indication for a (positive) correlation between stated skill and knowledge, we further looked into the course scores, consisting of graded multiple choice tests, programming assignments and a peer assessment yielding bonus points.
Calculating the average scores for the respective skill groups indicates a different distinction for these values: 11.4, 21.1, 30.4, 34.5, and 46.33 out of a maximum of 80 points.
The Kendall-correlation between the skill levels and the course scores turned out to be even lower ($0.18$) than that between the skill levels and the multiple-choice questions.
While we expected to gain a higher correlation for the more extensive course scores, it is likely that many participants compensated lesser prior knowledge with diligence and thus reached higher scores than to be ``expected'' from the self-stated skill level, leading to a lower correlation in our test.
As a last step, we calculated and plotted the completion ratio of the respective skill groups (see Figure \ref{pic:completePerSkillLevel}).
A participant is considered to have completed the course when at least 50\% of the points were reached, resulting in a minimum of 30 points for that course (60 points through exercises in total, up to 20 points bonus through the peer-assessment).

Written qualitative feedback in the forums concerning the exercises in general was positive, regardless of the difficulty of the exercises or the specific course.
This reassures us that the offering of practical exercises is valued, even if course instructors do not succeed in finding optimal suited exercises.

The individual feedback also uncovered potential pitfalls that are likely to distort the gathered metrics, albeit the actual problem lies not in the exercise: several beginners had problems finding the correct curly brackets used within the Java syntax on the keyboard.
This was not prevented by the first playground exercise. 
Although the playground exercise was intentionally ignored with concern to the metrics and gave the participants the possibility to accustom themselves with the development environment, its template already contained all uncommon characters, leading them to face this problem only later.
 
\section{Related Work}\label{ch:related-work}
This paper contributes to the research areas Knowledge Assessment and Learning Analytics in MOOCs.
Research in this field seeks to understand the effects of teaching and learning, especially within online learning environments.
While the effectiveness of learning is measured via many different models (such as Kirkpatrick's 4 levels of evaluation~\cite{kirkpatrick_evaluating_1994}, the ROI methodology by Kirkpatrick and Philipps~\cite{phillips_value_2007}, the Six Sigma approach~\cite{george_lean_2004, smith_six_sigma_1993} and others), these models do not identify practical means to conduct or improve the actual learning. 
Albeit they come with a detailed level of descriptions (reaction of participants to trainings, changes in job workflows, and in the end the return on investment as pure monetary value in the ROI methodology) or calculations of improvements for production processes (Six Sigma), they also have no measures for the initial knowledge. 
Additionally, these models were designed for business situations, meaning staff trainings, conducted most often in class room settings or even smaller scales.
For these reasons, suited approaches to determine a knowledge status quo have to be  designed another way.

\subsection{Self Assessment}
As any metric extracted from a questionnaire, the reliability of self-assessment depends on neutral wording and the offered choices, including the number of options to pick from \cite{jones_optimal_2013, behling_translating_2000}.
Furthermore it should be taken into account, that people tend to avoid the border options and might show overconfidence \cite{kruger_unskilled_1999}.

Models such as proposed by Raphael Poss~\cite{poss_how} are likely to be well suited for hiring situations. The developed matrix distinguishes several important skill areas and reflects the well established CEFR\footnote{Common European Framework of Reference for Languages} levels. The supplied questionaire of 104 questions to infer the most likely level is not suited for a MOOC context, as we argue that even just reading the different level descriptions in the matrix is too time consuming in our context.
In order to get reliable data, the idea of simply forcing participants to answer more extensive questionaries in order to advance in the course is not only against our goals and intents, but given current average completion rates over all analyzed MOOCs of around 15\%\footnote{Data retrieved from http://www.katyjordan.com/MOOCproject.html on 11th October 2016}, the risk of loosing participants is too high. Courses requiring higher involvement by offering a mixture of automated and peer grading show a lower completion rate of about 10\%. For courses relying only on peer-assessment, the completion rate is even lower, at 5\% on average.

When designing a course, the intended audience plays a major role with regard to students' expectations, tenacity and prior knowledge.
However, even when aiming just for beginners, chances are that many advanced participants will join the course, as they are interested in the topic discussed. 
This raises the danger that the beginners feel intimidated or overburdened and are thus hindered in leveraging the full potential of the course. 
This effect was also noticed in other courses on other platforms~\cite{liyanagunawardena2015who}.

\subsection{Assessment Based on Course Progress and Quizzes}

Google uses Skill Maps in their Course Builder Platform~\cite{roussev_course_2016}. They define a skill as a unit of knowledge taught in a course, which might be composed of videos, text or other activities. 
Between these skills there may exist relations, such as ''depends-on´´, ''follow-on´´ or others. This in turn allows participants to navigate through a course in the context of the skill-graph, instead of having to follow the mostly linear course preset.
While such a skill map is suitable to reflect the skills acquired within a course, the authors do not state on how to assess prior knowledge with it.
In our view, after an introductory self-assessment, it might be feasible to give students who stated they already possess knowledge about the covered topics more advanced exercises and relate their success to the depending basic concepts afterwards with such a skill map.

\subsection{Assessment Based on Assignment Evaluation}
As the topics of our courses were Java Programming, Python Programming and unit testing within Java, we could yield additional information that is only available when solving practical development tasks. Of particular interest might be required time, error rates and issued program runs within such a context.

Previous work has either analyzed this in classroom settings or rather small MOOCs~\cite{yudelson_investigating_2014}.
The data was used to form student models and reflect their knowledge acquisition.
The authors extracted concepts from the code and used the data as an example on how to automatically derive learning paths.
Their focus thus was not the actual prediction of students' performance in MOOCs.

The findings of Liyanagunawardena et al.~\cite{liyanagunawardena2015who} in a repeated game programming MOOC reflect those that we also encountered:
skill gaps can affect a course positively (sharing of knowledge) as well as negatively (''... course was hijacked by experienced programmers...´´).
The authors mostly rely on a user survey and do not further state whether they analyzed any programming specific metrics.

The research of Olsen et al.~\cite{olsen_predicting_2015} is not about programming specifically, but in an adjacent field. They created several models for predicting the success of pupils solving math problems collaboratively.
They specifically evaluated the effects of collaborative work.
While especially the aspect of collaboration is of interest also for programming in the long run, specifics on the exact metrics used to create their models can not directly be related to our field of programming, as the authors mainly used binary information whether a calculation step was performed correctly and derived learning rates from that.
Apart from that, the number of participants ($N=84$) is not comparable to a MOOC setting.

\subsection{Complex or Specific Models for Knowledge Assessment}

Ni{\v{z}}nan et al. use complex models to predict the potential score outcomes~\cite{rihak_student_2015}. 
Starting from the Elo rating system, extensions like bayesian modeling are used to improve the accuracy. 
Also hierarchical models are employed to distinguish more fine grained concepts instead of just representing a general skill level. 
While the extensions increased accuracy sometimes, the authors state the improvements were only small.
This encourages us not to start with too complex approaches. 
Furthermore, the authors state that there are domains which require a deeper understanding of the supplied material than just learning the mere facts.
For those domains, it may be more worthwhile to add extensions that cover the specific relations within the concepts to be learned.
  
A comprehensive overview of existing models can be found in Desmarais' and Baker's work~\cite{desmarais_review_2012}.
If the specific moment or timespan when a participant gained a particular skill is relevant, Baker proposes a machine-learned model to assess the probability that the skill was learned at a certain learning item~\cite{baker_detecting_2011}.

Apart from the mentioned approaches, it is also possible to incorporate other information for our assessment process, such as demographics like age, the highest degree, or results from previous courses, be them related or unrelated with the current question.
According to Morrison and Murphy-Hill, the age and skill of a programmer (proxied by stack overflow reputation in their study) show a positive correlation~\cite{morrison_2013_programming_age}. 
However, as we currently do not have thorough data on these demographics across all participants and we did not  want to ask our participants for such relatively private data just for the sake of assessment, we omitted these considerations from our study.

Combining the many different areas of general programming such as debugging, profiling, or structuring code to just a single number that will represent the skill in our case, is also potential pitfall.
Albeit there is likely more potential to assess and represent these knowledge areas within a complex model, boiling down the complexity to a single number enables us to take action based on the data much easier and to test our assumptions relatively early.
 
More advanced topics such as execution parallelization or hardware optimizations are not taken into account, as they are not likely to be discussed in beginner courses and will most likely be thought in a dedicated course on their own.
The whole field of domain knowledge can also excluded in our case, since we assume that MOOCs on learning how to program try to stay as abstract and general concerning their domain in order to ensure accessibility.

Previous work either analysed (self-) assessment in classroom settings or hiring situations.
While there is plenty of literature explaining potential factors of success or failure of courses and reached scores, approaches for a-priori analyses of knowledge in MOOC settings currently lack deeper insights.
This work presented concepts, first steps and claims within this area in order to start filling the current gap.


\section{Future Work}\label{ch:future-work}
 The basis of a successful programming course will always be a corpus of good introductory material (mostly videos) and carefully crafted exercises that have a direct relation to the taught concepts. 

With regards to the exercises, we will have a closer look at the influence of description texts towards the metrics (mainly score and working times).
If the cause for struggles is an exercise itself, as could be seen on the exercise on polymorphism, the exercise should be fixed or simplified as soon as possible to solve this problem.
We suggest triggers within the platform to automatically alert instructors if an exercise shows a deviation of average working times greater than for example 50\% of the average of all course exercises.
Since the expected difficulty of an exercise mostly relies on evaluated guesses of course instructors, we integrated a feedback mechanism into the platform that asks a share of the participants for their perceived difficulty level in order to constantly approximate the actual difficulty of an exercise for the audience.
If the cause for struggle is individual to a participant, we want to cover this with interventions, either to take a break, to search for additional information or to ask their fellow participants for comments on their code.
First, promising experiments have shown that user acceptance on the commenting approach is given, with a positive side effect that especially expert users are motivated to help out, as this poses them new challenges and yields encouraging feedback in terms of individual valuation and general reputation.
The evaluation of the effect of these interventions on scores and course results will be covered in a subsequent paper.

As users expressed having difficulties with different concepts, we will also try to detect their concept specific weaknesses by tagging the exercises with the applied concepts and thus will be able to relate their working times and scores with the respective concepts.
This will then allow us to propose bonus exercises suited for their needs, in order not to overwhelm them with unnecessary tasks but to supply additional helpful exercises.
As intriguing as the idea of a course consisting of completely individualized exercises with regards to the amount and the task difficulties may seem, we do currently not see this feasible with regards to the workload that would be imposed on the instructors.
Because this would also rise questions concerning fairness in grading, we therefore will treat individualized exercises as a field for experimentation on ungraded and bonus exercises solely.

 In order to improve the assessment of prior knowledge, the work of Leinonen et al.~\cite{leinonen_preventing_2017} hints that specific patterns such as curly brackets or particular common expressions like ``i++'' and their typing speeds allow to deduce users' skill levels to a certain degree.
 

 \section{Conclusion}\label{ch:conclusion}
 Programming exercises are a vital part of computer science MOOCs.
 Exercises with increasing difficulty and complexity seem to be accepted by wide audiences, as long as the average working times stay in a reasonable timeframe of 5 to 25 minutes each.
 Optimal exercises should challenge participants, but always be backed by suitable explanatory material.
 Even if a particular exercise is too demanding, participants forgive such a mismatch as they valuate the offered possibility for hands on practice over minor didactical shortcomings.
 Estimating the actual difficulty of assessment questions and programming tasks beforehand is hard for the course instructors, as the intended and actual audience might differ greatly.
 Furthermore, the effects of participants' different skill levels onto the required time and the problems to tackle are hard to pre-estimate.
  When participants are regularly exposed to exercises perceived as too difficult, they react by omitting them if a pattern is recognizable.
 In order to personalize learning experiences through bonus exercises that are suited towards individual weaknesses, deviations of individual working times and scores compared to the course averages can be used to determine the best exercise.
 With regard to exercise specific metrics such as the average score, the number of runs and required working times, it became apparent that these metrics do correlate with each other.
 Total reached scores did not turn out to be a practical measure to estimate prior knowledge if participants are allowed unlimited attempts, as determined users compensate missing knowledge with diligence.
 For an ad-hoc assessment of prior skills, self-classification of participants backed by a small number of multiple-choice questions yielded promising results towards estimating users' course success.
 A detailed analysis of working times and error messages shows greater potential for the assessment of prior knowledge during course runtime.
 In order to decouple the suitability of exercises from the knowledge of participants to some extent, we propose assistive tools for helping participants out through collaboration in the moment they are facing problems.
 Leveraging the very potential within MOOCs, their massive user base, looks promising for levelling out knowledge gaps and offering additional challenges for participants on expert level.
 Given a basis of exercises generally suited for most users, collaboration tools thus not only fill individual shortcomings, but also increase involvement and communication.
 Individual problems with exercises thus can serve as starting point for meaningful discussions.
 Therefore, a last provocative claim arises: the optimal practical programming exercise, at least for MOOCs, might therefore be a flawed one.

\bibliography{optimal-exercise}
\bibliographystyle{plain}

\end{document}